\documentclass[a4paper,10pt]{article}
\usepackage[utf8x]{inputenc}
\usepackage{amsmath,amssymb}
\usepackage{tikz}
\usetikzlibrary{3d,calc}
\usepackage{graphicx}
\graphicspath{ {images/} }
 \usepackage[toc,page]{appendix}
\usepackage{hyperref}

\newsavebox{\uuunit}
\sbox{\uuunit}
    {\setlength{\unitlength}{0.825em}
     \begin{picture}(0.6,0.7)
        \thinlines
        \put(0,0){\line(1,0){0.5}}
        \put(0.15,0){\line(0,1){0.7}}
        \put(0.35,0){\line(0,1){0.8}}
       \multiput(0.3,0.8)(-0.04,-0.02){10}{\rule{0.5pt}{0.5pt}}
     \end {picture}}

\newcommand{\ba}{\begin{eqnarray*}}
\newcommand{\ea}{\end{eqnarray*}}
\newcommand{\ban}{\begin{eqnarray}}
\newcommand{\ean}{\end{eqnarray}}

\def\beq{\begin{equation}}
\def\bee{\begin{equation}}
\def\eeq{\end{equation}}
\def\bea{\begin{eqnarray}}
\def\eea{\end{eqnarray}}
\def\bd{\begin{displaymath}}
\def\ed{\end{displaymath}}

\begin{document}

%opening
\title{Massive Anti-de Sitter  Gravity  from  String Theory \\
  \vskip 5mm
\,  }

\author{Constantin  Bachas and Ioannis Lavdas}

%\begin{document}

\maketitle
\begin{center}
\textit{Laboratoire de Physique Theorique de l'Ecole Normale Superi{\'e}ure,\\ 
PSL  University, CNRS  {\rm \&}  Sorbonne Universit\'es \\ 
24 rue Lhomond, 75231 Paris Cedex, France}
\end{center}

\vskip 1cm

\begin{abstract}

\end{abstract}
%\begin{center}
We  study  top-down embeddings of massive Anti-de Sitter (AdS) gravity in type-IIB string theory.
The supergravity solutions
%,  dual to the interface  conformal theories  of Gaiotto-Witten, 
 have a   AdS$_4$ fiber  warped over a
manifold M$_6$ whose shape resembles that of scottish bagpipes:  The `bag'  is  a  conventional  
AdS$_4$-compactification manifold, while the `pipes'  are  highly-curved  semi-infinite Janus throats. 
Besides  streamlining previous  discussions of the problem, our 
 main new result is   a   formula for the graviton mass which  only depends on the effective gravitational
coupling of the bag, and on the D3-brane charges   and  dilaton jumps  of  the Janus throats. 
We compare these embeddings   to the Karch-Randall model  and 
to other  bottom-up proposals for massive-AdS-gravity,  
and we comment on their   holographic interpretation.
This is a companion paper to   \cite{Bachas:2017rch}, 
where some  closely-related bimetric  models with pure AdS$_5\times$S$^5$ 
throats were analyzed.

%\end{center}
\newpage

%%%%%%%%%%%%%%%

\noindent{\bf\large 1.   Introduction}
\vskip 1mm

  %  Infrared modifications   of  Einstein's  gravity  are  of  both  theoretical and  observational interest.
     Efforts 
    to endow  the graviton  with a  tiny  mass have a long history,  going back to  the  work of 
   Fierz and  Pauli   \cite{Fierz:1939ix} and continuing   unabated today -- for  reviews and  references see
       \cite{Hinterbichler:2011tt}\cite{deRham:2014zqa}\cite{Schmidt-May:2015vnx}. 
   The problem  is of obvious theoretical interest,  and could
   have far-reaching  implications for cosmology. In one recent  development 
 it  has  been argued that a 
 key  obstruction to the  graviton mass  --  the appearance  of a   Boulware-Deser
   ghost \cite{Boulware:1973my},   can be removed  in certain  classical  non-linear  extensions of the Fierz-Pauli  action \cite{deRham:2010kj}\cite{Hassan:2011hr}(see also \cite{Chamseddine:2018qym}).
    Questions,  however,   remain  both in what concerns  the consistency of such  classical theories, and 
    with regards to their range of  validity if viewed  as effective  
   field theories around a  given classical   background.\,\footnote{For a very partial list of references  raising  or 
   trying  to address  such questions  see  \cite{ArkaniHamed:2002sp}\,-\,\cite{deRham:2018qqo}
   and the above reviews.
   %\cite{Chamseddine:2013lid}\cite{Deser:2013qza}\cite{Deser:2014fta}\cite{Hassan:2017ugh}
   %\cite{Bellazzini:2017fep}
  } 
          
                       One may  hope to
         answer such  questions by embedding massive gravity in an ultraviolet-complete theory like string theory. 
             In this paper we  will consider four-dimensional  Anti-de Sitter (AdS) solutions of IIB string theory  in which the  lowest spin-2 mode has   a tiny mass $m_{\rm g}$.
    In discussions of massive gravity,  AdS 
 is known to be an `easier  case'   since it   does not suffer from  some  of the difficulties encountered in 
 the   Mikowski and  de Sitter backgrounds. 
        There is, in particular,    no 
          van Dam-Veltman-Zakharov discontinuity  \cite{Kogan:2000uy}\cite{Porrati:2000cp},      
         and  hence no  need for the  strong non-linearities known as  Vainshtein screening 
           \cite{Vainshtein:1972sx}.  
      It remains therefore to be seen whether our  embeddings of massive AdS gravity  carry any
        lessons for these other backgrounds.

                 The general idea behind  the  embeddings, due to   Karch and Randall
                    \cite{Karch:2000ct}\cite{KR2}, is  to `locally localize'   the graviton  on 
                 an AdS$_4$  brane living   in  AdS$_5$ bulk. Using  a thin-brane approximation these authors 
                showed that if  the  ratio of  AdS  radii is small, 
                  $L_5/L_4\ll 1$,    the  lightest graviton mode acquires a  tiny mass.  The proper string-theory  realization 
                  of the  idea had to wait for the  derivation of  exact  solutions describing   intersecting D3\,, D5 and 
                  NS5-branes  \cite{D'Hoker:2007xy}\,-\,\cite{Assel:2012cj}\,. 
                  %\cite{D'Hoker:2007xz}\cite{Bachas:2011xa}\cite{Aharony:2011yc}
                  One  key departure from the Karch-Randall model
                    is the failure of the thin-brane approximation for  the localizing
                  brane,  which is  a  D3-D5-NS5  bound state. As shown in 
                   \cite{Bachas:2011xa} the  AdS radius of  this
                  composite  brane cannot be made parametrically larger than its thickness. As a result 
                    the
                  Kaluza-Klein scale  (beyond  which any  $4d$  description must break down)  is $L_4$,  and not
                   $ L_5$ as    in ref.\,\cite{KR2}. This is related to the familiar  scale separation
                  problem of AdS flux vacua, for a discussion see  
                  \cite{Polchinski:2009ch}\cite{Tsimpis:2012tu}
                   \cite{Gautason:2015tig}.\,\footnote{As explained in 
                  ref.\,\cite{Polchinski:2009ch} scale non-separation is unavoidable  for solutions with   a continuous
                  R-symmetry, i.e. for the vast  majority of   known solutions
                  of $10d$ supergravity including the ones discussed here. 
                  The authors of this reference outline  possible ways out of the problem.}

                        The purpose of the present note is to derive an (almost  universal)  formula for the graviton mass
                        in  these string-theory embeddings. This is a follow-up paper to ref.\,\cite{Bachas:2017rch}
                        which analyzed closely  related  embeddings of  
                        bigravity models.  Apart from the change in  emphasis compared to \cite{Bachas:2017rch},   
                        we will here  also  extend the results of  this reference 
                        by allowing   the dilaton to vary  in the AdS$_5$ bulk  which is  
                        deformed to    the  supersymmetric
                         Janus background
                         \cite{D'Hoker:2007xy}. This modifies  the graviton mass 
                         by a multiplicative factor that we will compute.
                         Our formula  for the graviton mass is derived  on the gravity side. It 
                     is an interesting open problem   
                         to match it   with  a computation of the  anomalous  dimension  of the energy-momentum tensor
                         on  the CFT side.

                   There have been two other proposals in the literature
                   for realizing massive AdS gravity  in string theory. They
                   relied    either on transparent boundary conditions in AdS 
                   \cite{Porrati:2001db}\cite{Duff:2004wh},  or on multi-trace
                   deformations in  CFT  
                    \cite{Kiritsis:2006hy}\cite{Aharony:2006hz}\cite{Kiritsis:2008at}.  In these proposals
                    the graviton mass is a quantum, one-loop effect. 
                            Although our embeddings  could be
                   possibly rephrased  in   these other frameworks by  integrating out messenger degrees of freedom,
                   they have the  advantage   of relying  on
                   proper  classical solutions of $10d$ supergravity. 
                    They  do not therefore suffer from  some  difficulties of  the above  proposals, namely
                    non-locality   of the worldsheet  theory,   or  hard-to-control renormalization group flows
                   \cite{Aharony:2001pa}\cite{Aharony:2005sh}\cite{Aharony:2015afa}. 
                    The  `price to pay'  is that the graviton mass  
                     is quantized and cannot be tuned continuously to zero. We will  return to this point  towards the end.

          % \medskip

                       This paper is organized as follows:  In section 2 we recall  why  defect 
                       or interface CFT  \cite{Karch:2000gx}\,-\,\cite{Erdmenger:2002ex} is the appropriate holographic
                       setup for  Higgsing the AdS graviton.  
                          Holographic duality is not crucial to   our later analysis,
                       but it provides useful insights  on the underlying  mechanism.   Section 3 explains   the
                       group  theory of the Higgsing, i.e. the recombination of representations   of the ${\cal N}=4$ 
                       superconformal algebra $\mathfrak{osp}(4\vert 4)$  which is the symmetry of the relevant  background
                        solutions. This section  can be skipped without affecting  the flow of the paper.
                            
                       Section 4   describes   the   qualitative characteristics of the supergravity solutions that 
                       lead to a small  graviton mass. 
                        These solutions consist of   AdS$_4$  fibers  warped over  six-dimensional
                       manifolds  with    the shape of scottish bagpipes. The `bag'  describes a standard 
                            AdS$_4$ compactification, while the non-compact `pipes'  are 
                            highly-curved   Janus  throats.  In section 5 
                              we calculate, following \cite{Bachas:2017rch}, 
                            the graviton mass to leading order in the throat-to-bag size, and 
                            show that  it   only depends on few  parameters of the 
                            solutions: the radius and dilaton variation in  each throat,  and the  effective gravitational
                            coupling of the bag solution.  This is  the main technical   result of the  paper.
                           To extract its physical significance we  reexpress it in three different   ways.
                            In section 6  we comment on the relation to bimetric  and  multi-trace
                            models, while section 7 contains some  concluding remarks. 
                                     Explicit  expressions 
                                     for the metric and dilaton of the `bagpipes'  solutions
                                     are collected  in the  
                                       appendix.

       %%%%%%%%%%%%%%       
      
         \vskip 0.7cm 
           \noindent{\bf\large  2. Mass as  holographic leakage}                                            
            \vskip 2mm

          We  begin  our discussion of massive  AdS gravity   from the 
           dual   CFT side. This     sheds  instructive  light   on the  Higgsing
           mechanism and motivates the construction of the dual supergravity solutions. 
           Recall that holographic duality 
        associates to any AdS$_{4}$ vacuum of string theory 
a  three-dimensional conformal field theory (CFT$_3$). 
The AdS$_4$
  graviton is mapped  to the energy-momentum tensor $T_{ab}$ of the CFT$_3$,
and the mass ($m_{\rm g}$)  of the former to the scaling dimension
 ($\Delta_{\rm g} $) of the latter via   \cite{Aharony:1999ti}
\bea\label{1}
m_{\rm g}^2\,  L^2_{4} = \Delta_{\rm g}  (\Delta_{\rm g}  -3) \ , 
%\simeq 3+ \epsilon \qquad {\rm if}\qquad \Delta = 3+ \epsilon
%\ \ {\rm with}\ \ \epsilon\ \ {\rm small} . 
\eea
 where  $L_{4}$ is the  AdS$_4$ radius.\,\footnote{In  warped compactifications both 
 $m_{\rm g}$ and $L_4$ vary in the transverse space, but their product is constant -- see below.}
The operator $T_{ab}$ and   its tower of derivatives arrange themselves 
 in a   spin-2 highest-weight  representation
  of the conformal algebra  $\mathfrak{so}(2,3)$. 
 % Following standard conventions we   denote this   representation  $D(\Delta_{\rm g} ;  2)$. 
Usually the  
   energy-momentum tensor  is conserved, 
         $\partial^b T_{ab}=0$, so  this  representation must be  short 
         since  it has   three null descendant states.  
A simple algebraic computation   then shows  that $T_{ab}$ must have  canonical scaling dimension 
$\Delta_{\rm g} =3$, 
and   the dual
AdS$_4$  graviton is  hence massless. 
\smallskip

To obtain a massive graviton we must therefore  allow  $3d$ 
  energy-momentum  to `leak out'.\,\footnote{Leakage of $T_{ab}$  should not be confused with the transparent boundary conditions of \cite{Porrati:2001db}  on the gravity side.} 
      There are two possibilities    that are consistent with 
  $\mathfrak{so}(2,3)$ symmetry: \\ \vskip -3.8mm
 \indent \hskip 0.2cm 
  (i)   Couple  the original theory to  another  $3d$  theory 
 so that conformal symmetry is preserved. The  \\ \indent \hskip 0.75cm
 coupling could be a double- 
  trace deformation  \cite{Kiritsis:2006hy}\cite{Aharony:2006hz}, 
  or it could be mediated  by  messenger    \\ \indent \hskip 0.75cm   degrees of freedom \cite{Bachas:2017rch}. 
    If it is   weak  the dual low-energy string theory is a
  bimetric  theory,   \\ \indent \hskip 0.75cm with one   graviton  massless and the other
  obtaining   a small mass; \\  \vskip -3.8mm
  \indent
   \hskip 0.2cm  (ii)  Consider  the  original theory  as a defect  or boundary  
   of some higher-dimensional theory, in  \\   \indent \hskip 0.8cm    the simplest case 
  a  CFT$_4$.  The  $3d$  energy-momentum can now
  leak out in the extra  dimension 
  \\   \indent \hskip 0.8cm   $\partial^b T_{ab} =  T_{a4}\vert_{\rm defect} \not= 0$. 
   There is  therefore  now no   shortening  condition,    and $T_{ab}$  acquires an    \\   \indent \hskip 0.8cm  
     anomalous dimension  \cite{Aharony:2003qf},  $\epsilon = \Delta_{\rm g} -3 >0$ (unitarity requires that it be non-negative).

\smallskip
 
 \noindent These two options are related -- we   will here focus on  option  (ii)  which can be 
obtained  as a  limit of  option (i).  Since  the graviton mass is proportional to 
 the $3d$ energy-momentum leakage, we want the latter to  be weak.\,\footnote{In generic  defect CFTs
 the defect-to-bulk leakage
is  very strong  and   $\Delta_{\rm g} - 3
 \sim O(1)$. In such cases  there is no
hope of any  effective $4d$  description on the gravity side.} 
 In principle this could be achieved by fine tuning
a  (nearly or exactly)  marginal  bulk-boundary coupling,  
but this is not the mechanism at work here.  
Weak leakage will be   instead ensured by the scarcity of the bulk CFT$_4$ degrees of freedom,  as compared to those
of  the boundary CFT$_3$.   A consequence of this is that   the Higgsing  
 will  not be a continuous process in these   models,  even though the graviton mass can be arbitrarily small.

         Let us be now  specific about the defect CFT.   The natural candidate for the  bulk CFT$_4$ is  ${\cal N}=4$ 
     super Yang-Mills with gauge group $SU(n)$ and coupling $g_{\rm YM}$.   Its 
     half -BPS superconformal boundaries and interfaces
     have been analyzed by Gaiotto and Witten  \cite{Gaiotto:2008sd}.  Half-maximal  supersymmetry
     guarantees the stability of the solutions,  and gives extra   technical control,   but it  is  not  otherwise
     essential.  The graviton mass, in particular,    is not a
     protected quantity as we will see  in a minute. 
    Weak leakage of   $3d$ energy-momentum   could be achieved in  the  decoupling 
    limit $g_{\rm YM}\to 0 $, but this limit is  singular.  A better alternative is to  insist that  there are much fewer    degrees of freedom in the bulk than on the boundary.  
     We will indeed show  in section 5  that the anomalous dimension of $T_{ab}$ 
     scales  like  $\epsilon \sim n^2/\tilde F_3$,    where 
       $\tilde F_3$   is the    free energy  on the 3-sphere
       which  measures  the boundary degrees of freedom \cite{Giombi:2014xxa}.

   %%%%%%%%%%%%%%       
      
         \vskip 0.7cm 
           \noindent{\bf\large  3. Recombination of representations}                                            
            \vskip 2mm

      Before moving to geometry, let us discuss the Higgsing from the point of view of representation theory. 
      Let $D(\Delta, s)$  denote a unitary highest-weight representation  of $\mathfrak{so}(2,3)$
      with conformal primary of spin $s$ and
    scaling dimension $\Delta$. 
     Massive gravitons  belong  to  long representations  of  the  algebra. 
     The  decomposition of a long spin-$s$ representation 
  at the unitarity threshold reads \cite{Porrati:2001db}
\bea\label{decomp}
D ( s+1+\epsilon ;  s)  \ \  
\overset{\small \epsilon \to  0}{\xrightarrow{\hspace*{1cm}}}    \ \  D(s+1 ;  s) \oplus D(s+2 ;  s-1)\ .  
\eea
Thus  the AdS$_4$  graviton  ($s=2$) acquires  a mass by  eating  a  massive Goldstone vector. 
    In the $10d$ supergravity this  vector must be  the  combination of   
 off-diagonal components  of the metric and tensor fields  that is dual to the
    CFT operator $T_{a4}$. 
    
      Since we will here deal with ${\cal N}=4$ backgrounds, fields and dual 
  operators    fit   in   representations  of the  larger superconformal 
   algebra $\mathfrak{osp}(4\vert 4)$. These  
  have been all classified  under mild assumptions
  \cite{Dolan:2008vc}\cite{Cordova:2016emh}.   In the notation of  
   \cite{Cordova:2016emh}  (slightly retouched   in  \cite{Bachas:2017wva}) 
the supersymmetric extension of the above  decomposition reads   \vskip -6mm
     \bea\label{3}
L[0]_{1+\epsilon} ^{(0;0)}   \ \
\overset{\small \epsilon \to  0} {\xrightarrow{\hspace*{1cm}}}  
%\longrightarrow \ 
 \ \ A_2[0]_1^{(0;0)} \oplus B_1[0]_2^{(1;1)}\ , 
 \eea
 where $L$ denotes  a long representation,  
 $A_i$\,($B_i$)   a  short 
 representation  that  is  marginally (absolutely)  protected, and $[s]_{\Delta}^{(j; j^\prime)}$ denotes a superconformal primary with 
 spin $s$, scaling dimension  $\Delta$ and  $\mathfrak{so}(4)$ 
  R-symmetry  quantum numbers $(j; j^\prime)$. 
  The above decomposition (or recombination) describes the Higgsing of the 
   ${\cal N}=4$ graviton multiplet  in  AdS$_4$. 
   That this   is at all possible  is not automatic. For instance 
 ${\cal N} =4$  supersymmetry  forbids  
the Higgsing of  ordinary gauge   symmetries   
because  conserved  vector currents  transform in  absolutely
protected representations of   $\mathfrak{osp}(4\vert 4)$ \cite{Louis:2014gxa}\cite{Cordova:2016xhm}.  
   
      The bosonic field content of the above  ${\cal N}=4$ multiplets is as follows:
\bea
  A_2[0]_1^{(0;0)} =  [0]_{1}^{(0;0)} \oplus [0]_{2}^{(0;0)}
\oplus
 [1]_2^{(1;0)  \oplus  (0;1) } \oplus
[2]_3^{(0;0)}  \oplus {\rm fermions}\ , 
\eea 
 \bea
 B_1[0]_2^{(1;1)} \hskip -2mm 
 &=& \hskip -2mm    [0]_{2}^{(1;1)} \oplus   [1]_{3}^{(1;1) \oplus  (1;0)  \oplus   (0; 1 )}
 \oplus   [0]_{3}^{(2;0)   \oplus   (0; 2 ) \oplus  (1;0)  \oplus   (0; 1 ) \oplus   (1;1 )
 \oplus   (0; 0 )}    \nonumber \\  [1em]
 &&   \oplus   [1]_{4}^{(0;0) \oplus  (1;0)  \oplus   (0; 1 )}  
  \oplus   [0]_{4}^{(1;1)   
 \oplus   (0; 0 )}  \oplus   [0]_{5}^{(0; 0 )}
   \oplus {\rm fermions}\ .   
\eea
  The supergraviton multiplet $A_2$ has in addition to the spin-2 boson,  
  six  vectors and two  scalar fields,  making  a total of  16 physical states.\,\footnote{The CFT operators
  include $T_{ab}$, six conserved R-symmetry currents and two scalar
  operators. Taking into account the conservation laws
  this gives also a total of  $25-9=16$ independent operators.}  The  eaten Goldstone multiplet $B_1$ contains
  112 physical bosonic states and as many fermions. These latter include massive spin-3/2 states which
  are not part of the spectrum   of gauged $4d$ supergravity \cite{Bachas:2017wva}. 
   Higgsing with that much supersymmetry is thus  necessarily a  higher dimensional process.

    %%%%%%%%%%%%%%       
      
         \vskip 0.9cm 
           \noindent{\bf\large 4.  Scottish bagpipes}                                            
            \vskip 2mm

         We turn now to the gravity side of the Higgsing. The local form of all  solutions of type-IIB supergravity
         with $\mathfrak{osp}(4\vert 4)$ symmetry has been derived by D'Hoker 
         {\it et al}  \cite{D'Hoker:2007xy}\cite{D'Hoker:2007xz}
         (see also  \cite{Lunin:2006xr}\cite{Gomis:2006cu}   for earlier work).  
         Global solutions  and the detailed  holographic dictionary have  been worked out in 
   \cite{Assel:2011xz}\cite{Assel:2012cj}\cite{Bachas:2017wva}. 
            All   solutions are  warped products of 
       AdS$_4$ over a base manifold M$_6$,  
         \bea
       ds^2_{10} \, =\,    L_4^{\,2} (  y)  \, ds^2_{{\rm AdS}_4}  +  \sum_{i,j=1}^6 g_{ij}({  y}) \, dy^i dy^j \ , 
       \eea
      where $ds^2_{{\rm AdS}_4}$  is   the metric of the unit-radius  Anti-de Sitter spacetime, 
      $ \{y^i \}$  are the coordinates of  M$_6$ with metric  $g_{ij}$,  and 
        $L_4(  y) $ is the   local radius of the AdS$_4$   fiber at a point $  y$. 
       The base manifold    M$_6$ is itself the warped product of two 2-spheres over   a
       Riemann surface.
       %, M$_6 \simeq (S_1^2\times S_2^2)\times_w \Sigma$.
               The  complete
       AdS$_4\times $S$^2\times \hat{\rm S}^{2}$ \, fiber  realizes 
       the $\mathfrak{so}(2,3)\times \mathfrak{so}(4) \subset \mathfrak{osp}(4\vert 4)$
        bosonic symmetry  of the backgrounds.  
      \smallskip
                  
\begin{figure}[h]
  \vskip -4mm
\centering 
%\vskip 0.2 cm
\includegraphics[width=.76\textwidth,scale=0.71,clip=true]{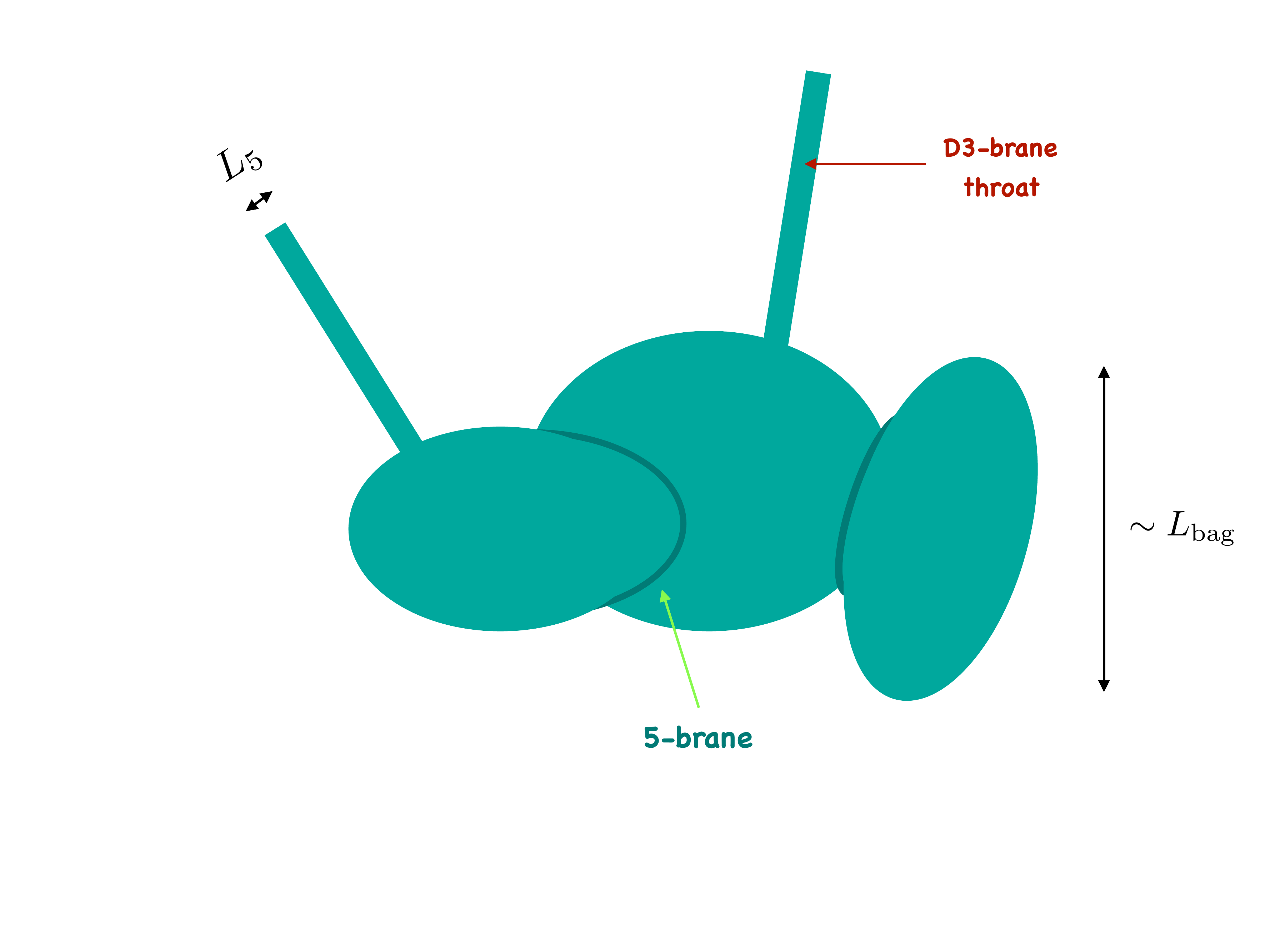}
   \vskip -16mm
   \caption{\footnotesize  The  `bagpipes'  manifold M$_6$  consists of 
   semiinfinite
   pipes  with cross-sectional  radius $L_5$,  attached to
   a compact bag  of typical size  $\sim L_{\rm bag}  \gg  L_5$. 
 The dark curves on the bag depict  5-brane singularities. 
 The AdS$_4$ scale  factor  diverges at infinity  in  the pipes so that the   full 10$d$  geometry  asymptotes to  (AdS$_5/\mathbb{Z}_2)\times$S$^5$.}
\label{fig:2} 
 \end{figure}

                 For  the lightest $4d$ graviton  to  acquire  mass,                        
              M$_6$   should not be a compact  manifold. The manifolds that 
                lead to a small  graviton mass  actually  resemble  
   six-dimensional Scottish  bagpipes: they 
      have one or more  semi-infinite throats  (the `pipes')  attached to 
     a large central core (the `bag') as illustrated   in figure \ref{fig:2}\,.  
     The full  $10d$  geometry of the 
     pipes is  AdS$_5\times $S$^5$,   or its 
      Janus
       generalization \cite{D'Hoker:2007xz} in which   the dilaton is also allowed to vary. 
       A crucial  technical remark   \cite{Aharony:2011yc}\cite{Assel:2011xz} is that  under certain mild  conditions 
        (existence of both NS5-brane  and D5-brane
       charges in the bag)  pipes  can be shrank smoothly away\,\footnote{Strictly-speaking this
        is possible in the supergravity approximation. 
        In  string theory the  
       throat radii are quantized.}\, 
       
        % \begin{figure}[b!]

      \noindent  leaving behind simple   coordinate
       singularities. Doing this   reduces the bag to   a compact manifold  $\overline {\rm M}_6$,  and AdS$_4\times_w  \overline {\rm M}_6$ becomes  a standard AdS$_4$
       vacuum with a massless $4d$ graviton. 
       Since we want   the graviton to acquire a small mass, we   keep the 
         throat radii  finite  but much smaller than the characteristic  bag size. 
 
     The exact metric and dilaton backgrounds of the  solutions are summarized in the appendix.\,\footnote{The
     supergravity solutions are actually singular  at certain 2-cycles of M$_6$ that are 
     wrapped by D5- and NS5-branes. In their vicinity  higher-order stringy and loop corrections  cannot be neglected,
 but fortunately they    do not   affect the  computation of
     the graviton mass at leading order in $L_5/L_{\rm bag}$.}
      The bag  depends
     on a set of   integer D5-, NS5-  and D3-brane charges
      which can be arranged in two Young tableaux \cite{Assel:2011xz}. Most  of these
      will play however no role here. The only   relevant bag parameters   are (i) an overall  measure of
      its size $\sim L_{\rm bag}$ to be defined below, and (ii)   the values of the dilaton
       at the entries of the throats.   Note that the bag 
      is a sort  of   `composite  Karch-Randall brane'. Tuning the available parameters  to make it flat,
      as in ref.\,\cite{Karch:2000ct},  
        makes it so thick   that it ends up occupying  (figuratively, not literally) most of space \cite{Bachas:2011xa}. 
      As stated above this is a  facet of  the scale non-separation problem of AdS vacua:   $L_4$ 
      is parametrically tied  to the
      characteristic size of  $\overline {\rm M}_6$. 
 \smallskip

     The   spectral problem for spin-2 excitations around any AdS$_4$ supergravity solution
    was   set up  
      in ref.\,\cite{Bachas:2011xa} (generalizing  mildly \cite{Csaki:2000fc}). 
  Interestingly this  problem   only depends on the Einstein-frame metric, 
      and not  on  the scalar and flux background   fields.  Mass eigenstates
       factorize as  $\psi(  y)\, \chi_{\mu\nu}$,   where $\chi_{\mu\nu}$
        is  an  eigenfunction of the  wave operator in AdS$_4$, i.e. 
         ${\cal L}_{(2)}^{\rm AdS}\chi_{\mu\nu} = \lambda
           \chi_{\mu\nu}$ where  ${\cal L}_{(2)}^{\rm AdS}$ is the  (Lichnerowicz-Laplace)  operator     
            that   acts  on spin-2 (transverse-traceless)  excitations, and the eigenvalue 
            $\lambda$ is related to the mass via 
           \bea
           \lambda + 2 = m^2(y) L_4^{\,2}(y)\ . 
            \eea 
            Note that both the mass and the AdS$_4$  radius may vary as functions of the coordinates $y^i$,
            but their   product is constant. It is this invariant squared mass  that replaces the left-hand side
            of  eq.\,\eqref{1} in   warped  (as opposed to  direct-product) solutions. 
     
       The  Kaluza-Klein mass spectrum   is determined by the elliptic  operator  
       acting on the wavefunctions $\psi(y)$ on M$_6$ \cite{Bachas:2011xa} \vskip -6mm
      \bea\label{operator}
         {\cal M}^2 \psi := 
           -{ L_4^{-2} 
           \over \sqrt{g}} \,  \partial_i  \left(  L_4^{\,4} \sqrt{g}g^{ij}\,\partial_j \, \psi \right)  = (\lambda + 2) \,\psi \ . 
         %   \hskip 8mm {\rm where}\hskip 8mm    \lambda + 2 = m^2(\vec y) f^2(\vec y)
                  \eea       
     For  direct-product solutions with constant  $L_4$,  ${\cal M}^2 $
      is simply  the Laplace-Beltrami  operator  on M$_6$. 
   To  define the spectral problem we need also to provide  a norm. 
        With  canonically-normalized   fields in ten dimensions\,\footnote{In our convention  $\psi$ has dimensions
         of  the $10d$   gravitational coupling 
         $\kappa_{10} \sim  [{\rm mass}]^4$. An  overall multiplicative constant in  the norm 
         is   irrelevant as it   drops 
         out from the  expression for the mass.}
          the  Kaluza-Klein  reduction of the inner product  reads  \cite{Bachas:2011xa}
           \bea\label{6}
                     \langle \psi_1\vert \psi_2\rangle = \int_{{\rm M}_6}d^6y\,
                      \sqrt{g} \, L_4^{\,2}   \psi_1^* \psi_2\ . 
                     \eea
       The mass-squared  operator 
                               \eqref{operator} is thus hermitean and non-negative,  
                                 as expected. 
     
          To summarize this section,  we are interested in  the smallest eigenvalue of 
          the above mass operator,  ${\cal M}^2$,   for  
     manifolds    consisting of  a large 
       compact bag  ($\overline {\rm M}_6$)   attached  to one or more   thin  semi-infinite Janus throats.  
       It actually  turns out that   the    solutions  studied here,  
       whose CFT  duals are ${\cal N}=4$  linear-quiver gauge theories,  
         admit at most   two semi-infinite throats.  But more general 
          backgrounds based on star-like quivers could have more throats. 
    As will be clear,   each throat    makes   a separate contribution to  the squared mass of the graviton  in the 
    $L_5/L_{\rm bag}\ll 1 $ limit.

  \vfil\eject                   
       
      %%%%%%%%%%%%%%       
      
         \vskip 0.7cm 
           \noindent{\bf\large  5. Mass from Janus throats}                                            
            \vskip 2mm

               The general  spectral problem defined in  \eqref{operator} and \eqref{6} is a difficult one. 
               But we  are only 
                interested in  the smallest eigenvalue  given by    the  equivalent  minimization   problem 
                     \bea\label{minimize}    
                           \    \lambda_{0} + 2
                             =\ 
                             {\rm \bf min}_\psi \left[ \int _{{\rm M}_6} d^6y\,
                              \sqrt{g}\, L_4^{\,4}  \,  (g^{ij} \partial_i \psi^\star \partial_j\psi)   \right]
   \ \   \  \ {\rm  with } 
   \ \ \ \  \   \int_{{\rm M}_6} d^6y\, \sqrt{g}\,  L_4^{\,2}\, \vert \psi\vert^2 = 1 \ . 
   \eea         
     Here $\lambda_{0} + 2 = \Delta_{\rm g} (\Delta_{\rm g} -3)$   is the lowest eigenvalue of 
     ${\cal M}^2$,  and the expression in square brackets is  $\langle \psi\vert {\cal M}^2
                           \vert \psi \rangle $ after an  integration by parts.   If                    
                                    M$_6$ were replaced by the  compact bag   $\overline {\rm M}_6$ 
         (obtained by truncating the pipes)  the  minimum
         would have been  the  constant wavefunction
     \bea\label{C}
       \psi_0( y)  =   \left( \int_{\overline {\rm M}_6} d^6y\, \sqrt{g} \, L_4^{\,2}  \right)^{-1/2}\ := 
        \psi_{\rm bag}   \  
      % \ \ \ \ a^{-2} = <f^2>_{\rm bag} Vol_{\rm bag}
       \hskip 8mm
      \eea
        corresponding to a massless graviton.   But the  infinite pipes make  the 
       constant $\psi$   non-normalizable.  Indeed, 
       $ \sqrt{g}\, L_4^{\,2} $ reaches a minimum  value   $L_5^{\,8}$
      inside   the pipes,    then  blows up  at infinity where the $10d$ geometry asymptotes to that of  (half)
      the boundary of AdS$_5$. This  is explained in the appendix and
        illustrated in   figure  \ref{fig:3}\,.  
          Normalizable  wavefunctions  should  therefore go to zero   inside  the pipes.
        Furthermore,  it is clear from eq.\,\eqref{minimize} that   in order  to minimize the mass  $\psi$ should go to zero in the region of minimal
       $\sqrt{g} L _4^{\,4}$ where gradients  have lower  cost, as shown in   fig.\,\,\ref{fig:3}\,.                                     
 
  \smallskip 
  To make the argument precise,  separate the manifold M$_6$ in three parts:  ({\footnotesize  I}) the bag, 
        ({\footnotesize  II}) the 
         infinite throats, and ({\footnotesize  III}) the matching regions where  throats are
        attached to the bag.   Minimizing  ${\cal M}^2$ in region {\footnotesize  (I)}  sets  $\psi$ to a constant, 
  so  the bag does not  contribute to the graviton mass.  On the other  
  
  \medskip

          % \begin{figure}[b!]
        \begin{figure}[h]
  %\vskip -6mm
\centering 
%\vskip 0.2 cm
\includegraphics[width=.76\textwidth,scale=0.80,clip=true]{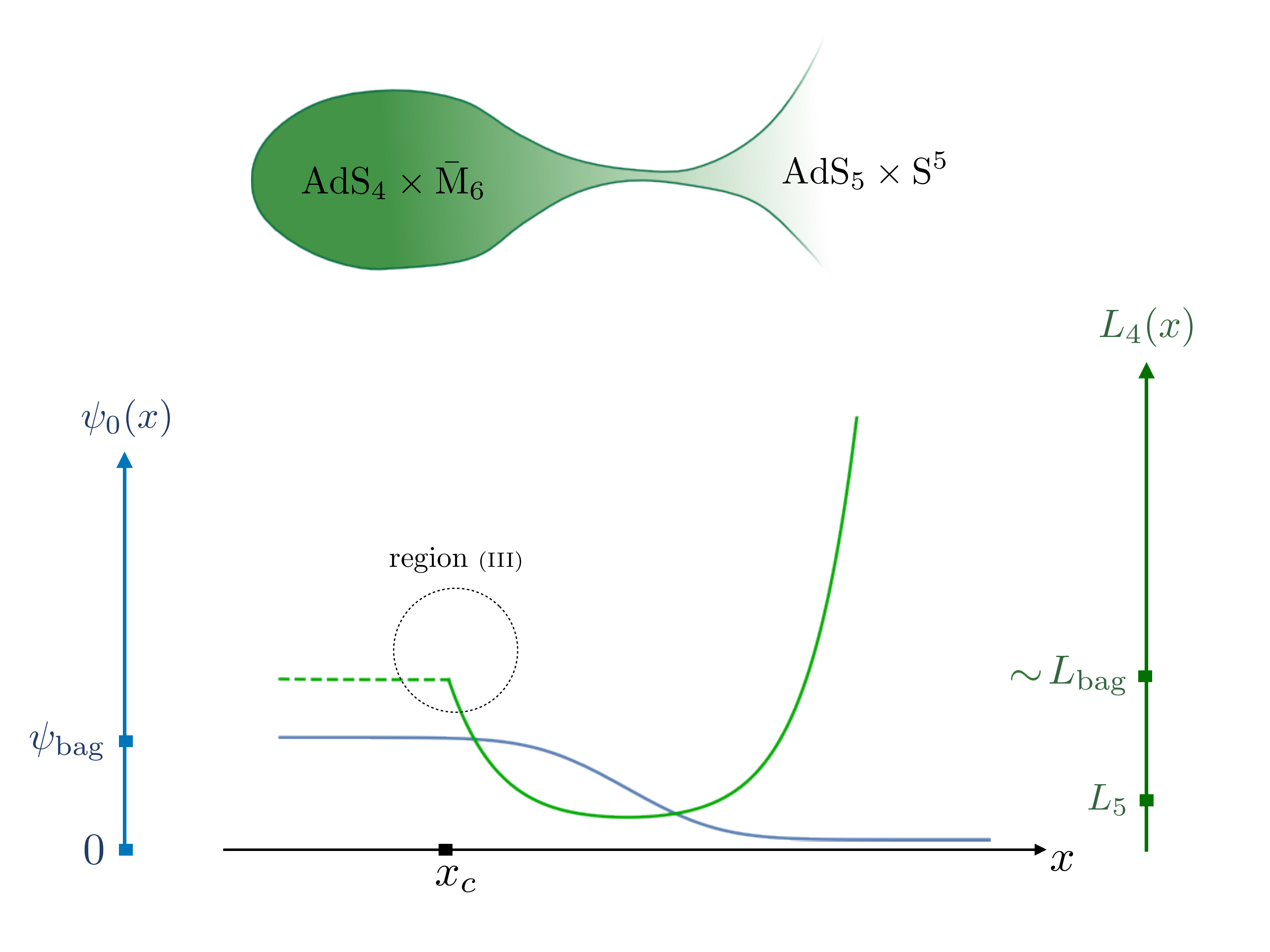}
  % \vskip -16mm
   \caption{\footnotesize Schematic drawing of the AdS$_4$ radius $L_4 $ (green curve)    and of the
   graviton wavefunction $\psi_0$ (blue curve)   as functions
   of the coordinate $x$ of the Janus throat.  The   radius  reaches a minimum value of 
   $L_5 $  inside the throat, and  grows  as 
   $L_5 \cosh x $ on either side 
     where the  geometry asymptotes to AdS$_5/\mathbb{Z}_2\times$S$^5$. 
    The  left half  of the throat  is 
   cut-off by the  `bag'   at
   a  characteristic   radius  $\sim L_{\rm bag}  \gg  L_5 $. The  graviton wavefunction  
   approaches a constant in this region, and 
   vanishes
   exponentially at   infinity. 
   Since the  matching region (III)  contributes  neither to the norm nor to  the mass, it
   can  be  shrunk for our purposes   to a point.
  }
\label{fig:3} 
 \end{figure}

        \noindent      hand, at  leading order in  $L_5/L_{\rm bag}$ 
          only the bag contributes to the norm of  $\psi$.  
     The reason is that 
           $\sqrt{g} L _4^{\,2}$ decreases exponentially fast in  the matching region,   and $\psi$
        vanishes exponentially fast in the throat as will be shown  in a minute. 
              This fixes   the constant value $\psi_0 \simeq  \psi_{\rm bag} $ in region {\footnotesize (I)}. 
              
          The region of minimal  $\sqrt{g} L _4^{\,4}$, on the other hand, where $\psi_0$ 
          can vary with minimal mass cost, lies deep inside the throat regions. Only the throats
           will therefore contribute to the graviton mass
        at this leading order, an 
         assumption whose validity 
        will be again verified a posteriori.

     In  summary,  the leading-order  contribution to the norm comes from the bag,   while 
     the leading-order  contribution to the mass comes from the bottom of the throat where $\sqrt{g} L _4^{\,4}$
     is minimal. The matching region ({\footnotesize  III}) contributes to
     neither and can be neglected.  We may thus  reformulate the problem  as 
     a variational problem in the Janus geometry: 
               \bea\label{minimize2}    
                           \   \lambda_0 + 2
                              \simeq \ 
                             {\rm\bf  min}_\psi  \left[ \int _{{\rm throats}}
                             \hskip -3mm d^6y\, \sqrt{g}\,  L_4^{\,4}  g^{ij} \partial_i \psi^\star \partial_j\psi  \right]
   \ \   \  \ {\rm with } 
   \ \ \    \psi  \to \ 
\begin{cases}
\psi_{\rm bag} \ \ \ {\rm in\ matching \ region },\\
\ \ 0 \ \ \ \ \ \ {\rm at \    infinity}\ .   
\end{cases}   
     \eea                      
The only residual dependence on  M$_6$ ({viz.}  on   the composite
NS5-D5-D3 brane)   is via  the boundary value  $\psi_{\rm bag}$  
 whose   physical meaning will  soon be made  clear.

 The  ${\cal N}=4$ supersymmetric Janus solution \cite{D'Hoker:2007xz} depends on    two parameters,  
 the radius $L_5$  and the dilaton variation $\delta\phi$. 
        Like all other solutions in this class it  has the fibered form
 \bea
  ds^2_{10} = L_4^{\, 2} ds^2_{{\rm AdS}_4} + ds^2_{{\rm M}_6} \qquad  {\rm with} \qquad 
  ds^2_{{\rm M}_6} = f ^{2} ds^2_{{\rm S}^2 } + \hat f ^{2} ds^2_{\hat {\rm S}^2} + 4\rho^2 dz d\bar z\ . 
  \eea  
  The scale factors  $L_4, f , \hat f, \rho$ depend on the  complex  coordinate  $z$ that  parametrizes  the infinite
strip. We write $z=x+i\tau $ with $\tau\in [0, \pi/2]$. The metric factors and dilaton are given in the appendix.
     Here we only need the combination that enters in the square brackets in \eqref{minimize2}.     
   Things simplify actually further because the  spin-2 eigenfunctions in the Janus geometry   factorize 
    into spherical harmonics on the  2-spheres,  and
  separate functions of $x$ and $\tau$, and 
       the  lightest  mode  is only function of  $x$ \cite{Bachas:2011xa}. 
       Integrating  over the 2-spheres and $\tau$ gives  \footnote{When 
$\delta\phi =0$ the  2-spheres and the coordinate $\tau$  make up the 5-sphere in  AdS$_5\times$S$^5$, 
and ${\cal G}(x) = 4\cosh^4x$.  } 
           \bea\label{x}
                     \lambda_0  +2 \ = \ {\rm\bf min}_\psi \left[
                        {\pi^3 \over 4} L_5^{ \, 8}   
                         \int_{x_c}^\infty \hskip -1.5mm dx\,    {\cal G}(x) \left( {d\psi\over dx} \right) ^{2}\, \right]  \  
                          \ \   \  \ {\rm with} 
   \ \ \    \psi(x)  \to 
\begin{cases}
\psi_{\rm bag} \ \ \ {\rm at} \ x= x_c ,\\
\ 0 \ \ \ \ \ {\rm at }  \  x=\infty\ , 
\end{cases}
\eea
where    the function ${\cal G}(x)$,  computed  in  the appendix,  reads
\bea\label{G}
   {\cal G}(x) :=  \left( { \cosh 2x + \cosh\delta\phi  \over \cosh\delta\phi }\right)^2 \    .
\eea                                                                                                                                                                                                                                                                                    
   We have  cutoff  the integral at some large negative value $x_c$,  at the boundary
of the matching region. The value of $x_c$ will drop out and could be replaced by $-\infty$, its only
role is to remind us that $\psi$ would have been a non-normalizable mode in the  complete  Janus geometry.       
         
             The   variational problem  \eqref{x} can  be easily solved,  
   \bea\label{vareqn}
                                                                                                                                                                                                                                                                                     {d\over dx} ({\cal G}  {d\psi_0\over dx}) = 0 \ \ \Longrightarrow\ \  \psi_0(x) 
                                                                                                                                                                                                                                                                                       =  c_1  +  c_2 \int_0^x {dx^\prime \over   {\cal G}(x^\prime)  } 
                                                                                                                                                                                                                                                                                   \eea
                                                                                                                                                                                                                                                                                    where   $c_1, c_2$ are  integration constants. 
                                                                                                                                                                                                                                                                                      We 
                                                                                                                                                                                                                                                                                     can  perform the integral  analytically 
                                                                                                                                                                                                                                                                                    with the  result  
$$                                                                                                                                                                                                                                                                                      \, \hskip -4cm  I(x,a) := \int_0^x { a^2\,  dx^\prime  \over (\cosh 2x^\prime  + a)^2}  = $$ \vskip -4mm
                                                                                                                                                                                                                                                                                  \bea =     { a^3   \over  2 (a^2-1)^{3/2}}\,\log \left[  {\sqrt{a+1 } +  \sqrt{a-1}\, \tanh x  \over  
                                                                                                                                                                                                                                                                                     \sqrt{a+1 }-   \sqrt{a-1}\,  \tanh x }   \right] -  { a^2  \over (a^2-1) } \, {  \tanh x\over
                                                                                                                                                                                                                                                                                      [ (a+1)  - (a-1)\tanh^2x \,] }\ \  .    \label{15}
                                                                                                                                                                                                                                                                                    \eea  
                                                                                                                                                                                                                                                                                    We here set $\cosh\delta\phi = a$ and chose  the  lower  integration limit 
                                                                                                                                                                                                                                                                                      so   that  $I$ is an odd
                                                                                                                                                                                                                                                                                    function of $x$.                                                                                                                                                                                                                                                                                     
     Fixing  $c_1$,$c_2$   so as to satisfy  the boundary conditions  \eqref{x} gives
     finally  the graviton wavefunction in the throat 
                                                                                                                                                                                                                                                                                     \bea \psi_0(x,a) \ \simeq \  {1\over 2}\psi_{\rm bag} \left[ 1 -  { I(x,a)\over I( \infty,a)}  \right]
                                                                                                                                                                                                                                                                                     \ .  \eea
                                                                                                                                                                                                                                                                                     Note that  $I$ approaches its limiting  values exponentially,  
                                                                                                                                                                                                                                                                                     so  $\psi_0(x_c) \simeq \psi_{\rm bag}$  up to exponentially
                                                                                                                                                                                                                                                                                     small  corrections. Furthermore at $x\to+\infty$, 
                                                                                                                                                                                                                                                                                      $\psi_0 = O(e^{-2  x })$  
 as required for  the norm to be finite.  The reader can now check that this  contribution to the norm  is 
 parametrically smaller than  that   of the bag,   and can  
 be  neglected as  claimed earlier. 

 Plugging the above wavefunction in the expression  \eqref{x} leads to  the
  graviton mass.  
 Note that    $\psi_0$ obeys eq.\,\eqref{vareqn}, so   the integrand is a total derivative and  one finds 
\bea \lambda_0  +2 \ = \  
                         {\pi^3\over 4}   L_5^{ \, 8}
                         \int_{x_c}^\infty \hskip -1.5mm dx\,    {\cal G}  \left( {d\psi_0 \over dx}\right ) ^{2}\,  =\, 
                          {\pi^3\over 4}   L_5^{ \, 8}  \left[ {\cal G}\, {d\psi_0 \over dx}\psi_0  \right] _{x_c}^\infty\ . 
                                                                                                                                                                                                                                                                                   \label{17a}  \eea 
                                                                                                                                                                                                                                                                                   \vskip 2mm
\noindent Since   ${\cal G}(x)   {d\psi_0 / dx} = - \psi_{\rm bag}/ 2 I( \infty,a) $  and 
 $[\psi_0]_{x_c}^\infty = -\psi_{\rm bag}$, we finally get 
 \bea\label{result}
                                                                                                                                                                                                                                                                                                                                                                                                                                                                                                                                                                                  \lambda_0  +2 \ =   \  
                         {\pi^3\over 8}   L_5^{ \, 8}\, \psi_{\rm bag}^2\, /  I(\infty, a) 
                         \ =\  {3\pi^3\over 4}   L_5^{ \, 8}\,  \psi_{\rm bag}^2 \,  J(a)\ , 
                          \eea
                                                                                                                                                                                                                                                                                            where we have introduced   the  Janus correction factor  $J(a)$, 
                                                                                                                                                                                                                                                                                          \bea\label{J}
                                                                                                                                                                                                                                                                                           J(a)^{-1}  := 6 \, I(\infty, a) =  {{3 }   a^3    \over  (a^2-1)^{3/2}}\,\log \left[ a + \sqrt{a^2-1}   \right] -  {3  a^2    \over  (a^2-1) }  \ . 
                                                                                                                                                                                                                                                                                          \eea

\begin{figure}[h]
   \vskip -4.2mm
\centering 
%\vskip 0.2 cm
\includegraphics[width=.61\textwidth,scale=0.65,clip=true]{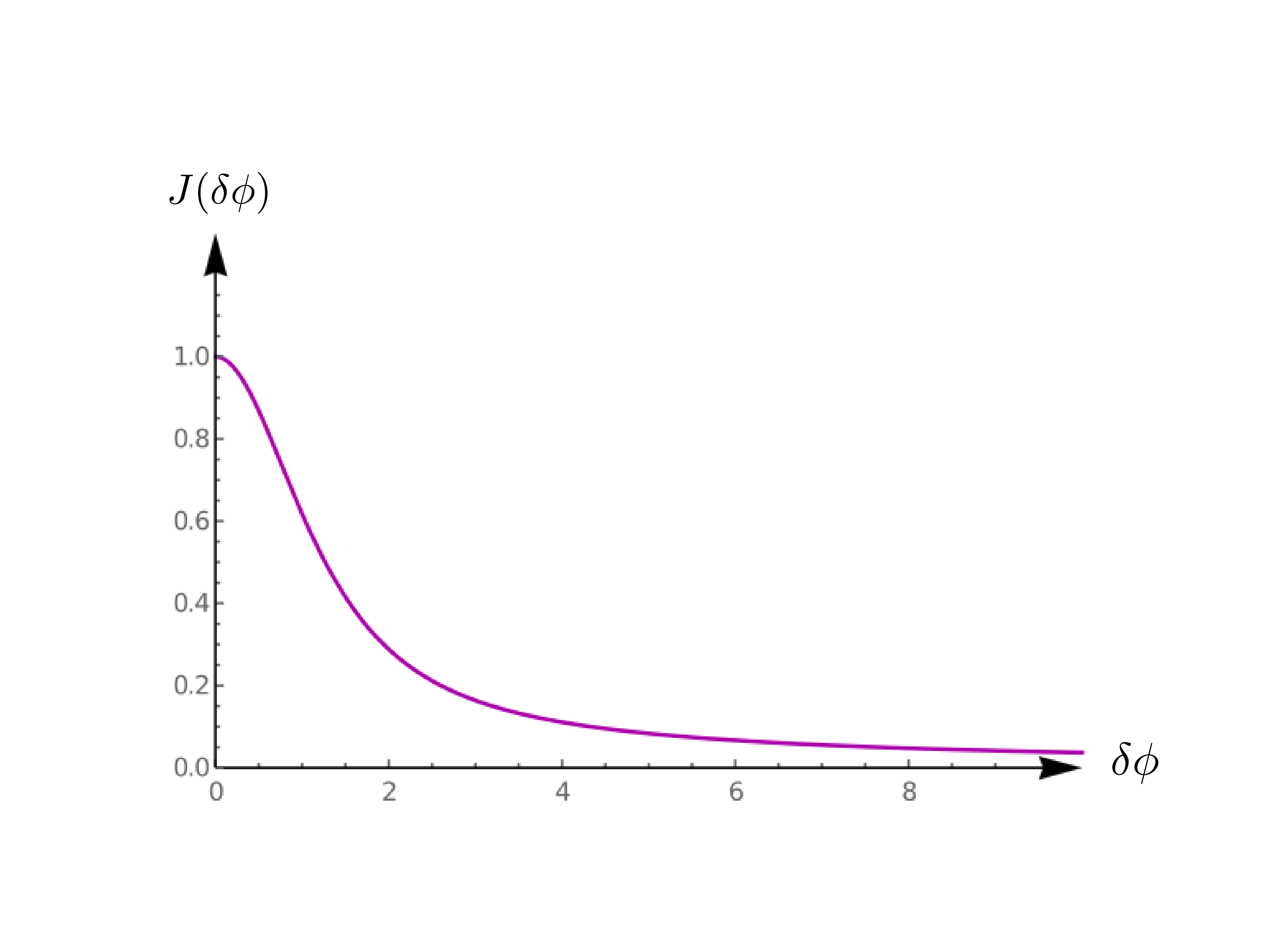}
    \vskip -12mm
   \caption{\footnotesize  The Janus correction function,  $J$,  defined in  equation \eqref{J}. 
  }   
\label{fig:4} 
 \end{figure}

                                                                                                                                                                                                                                                                                          As  $a = \cosh\delta\phi$ ranges from 1 to $\infty$, 
                                                                                                                                                                                                                                                                                           $J(a)$ decreases monotonically from 1 to 0,  see  figure  
                                                                                                                                                                                                                                                                                          \ref{fig:4}\,. The function  is normalized so as to  drop  out for  
  AdS$_5\times$S$^5$ throats, while  more generally it has the effect of reducing  the graviton mass.
  
                                                                                                                                                                                                                                                                                         This can be understood intuitively  as follows: 
                                                                                                                                                                                                                                                                                          $\delta\phi$  is the difference between  the value of the dilaton  at the entry
       of the throat and its value  at infinity.   The former is  fixed by the  bag 
    (see the appendix). The latter is a free parameter that 
   determines the  coupling constant $g_{\rm YM}$  of the dual  $4d$, ${\cal N}=4$ super Yang-Mills theory. Taking 
   $g_{\rm YM}$ to zero  (and hence $\vert\delta\phi\vert \to  \infty$)
   decouples the bulk  CFT  from the   defect,  restores conservation of $T_{ab}$ 
    and sends the graviton mass to zero.
                                                                                                                                                                                                                                                                                                                                                                                                                                                                                                                                                                                   The same is true,  by S-duality,   if $g_{\rm YM}$ is taken to infinity -- it is the  bulk magnetic theory now 
                                                                                                                                                                                                                                                                                                                                                                                                                                                                                                                                                                                   that decouples manifestly.  These limits are however singular. Not only does  supergravity  break 
                                                                                                                                                                                                                                                                                          eventually down, but also   the spectrum in the Janus throat  becomes quasi-continuous  \cite{Bachas:2011xa}
                                                                                                                                                                                                                                                                                          invalidating  any effective $4d$ description.

           % \begin{figure}[b!]

                                                                                                                                                                                                                                                                                          \vfil\eject

    Eqns.\,\eqref{result},\,\eqref{J}  are  the  main result of this paper.
   To extract their  physical meaning we will  rewrite them   in three different ways. 
   We first express  $\psi_{\rm bag}$,   
    defined in eq.\,\eqref{C}, in terms of  geometric data, 
   \bea
   \psi_{\rm bag}^{-2}\ =\ 
   \int_{\bar{\rm M}_6} \sqrt{g}\, L_4^2  \ = \  {  V}_{6}\, \langle L_4^2 \rangle_{\rm bag} \  
   \eea
     where 
     ${ V}_{6}$ is the volume of the bag,   and $\langle L_4^2 \rangle_{\rm bag}$
       the average AdS$_4$  squared radius  (which is finite 
       after truncating away the throats).  The 
     contribution  \eqref{result}   to the graviton
     mass thus  reads\,\footnote{If there are more than one  throats, one adds up  their contributions on the
     right-hand-side.}  
     \bea
 \underline{\rm geometric}:\quad \qquad      m^2_{\rm g}\, L_4^2 
                         \ =\    {3\pi^3  L_5^{ \, 8}\over  4 {  V}_{6} \langle L_4^2 \rangle_{\rm bag} } \times      J(\cosh\delta\phi)\ .  
     \eea
In the conventional  AdS$_4\times \overline M_{\rm 6}$     vacua the AdS radius is of the same order
as the Kaluza-Klein scale (this is again the scale non-separation problem), so  
  $(V_6)^{1/3}  \sim L_{\rm bag}^2 \sim \langle L_4^2 \rangle_{\rm bag} $.   Thus $m^2_{\rm g}$ 
       is  suppressed by the {eighth}  power of $L_5/L_4$, as  compared to 
      the second power which is the result 
         in the thin-brane
       Karch-Randall model \cite{Miemiec:2000eq}.     
  \smallskip  

     For a different   rewriting, we   use the relation between the  compactification volume  
     and the $4d$ effective  gravitational coupling,   $\kappa_{10}^2 = V_6 \, \kappa_4^2$\, 
     where $\kappa_{10}$ is the  coupling in ten dimensions.   Using    also the  relations 
     $2 \kappa_{10}^2 = (2\pi)^7 \alpha^{\prime\ 4} \lambda_s^2\ $  and $L_5^4 =  4\pi \alpha^{\prime\ 2}  \lambda_s\, n $, where  $\alpha^\prime$  is the Regge slope, $\lambda_s$ the string coupling constant
    (which can be absorbed in the definition of the Einstein-frame metric)
      and  $n$ the quantized D3-brane charge of the throat\,\cite{Polchinski:1998rr} we  get
        \bea\label{22}
   \underline{\rm gravitational}:\quad \qquad        m^2_{\rm g}\, L_4^2 
                         \ =\  {3 \, n^2   \over 16\pi^2  }
                         {  \kappa_4^2 \over  \langle L_4^2 \rangle_{\rm bag} } 
                           \times     J(\cosh\delta\phi)\ .  
     \eea
 This rewriting  in 
   terms of four-dimensional  parameters  brings forth two important points. First,   the  graviton mass  
    is quantized so  the Higgsing cannot be continuous \cite{Bachas:2017rch}. 
    Of course, to trust the supergravity
    approximation we need $n\gg 1$. The  mass does vanish with  two  continuous parameters,  $\kappa_4$ and
    $(\cosh\delta\phi)^{-1} $,   which can be taken to zero, but these limits are singular. 
    
    The second remark is that  \eqref{22}
    has the same  parametric form as the result  found   in refs.\,\cite{Porrati:2001db}\cite{Duff:2004wh}. 
     In these references the mass is a quantum effect  arising  from 
     integrating out matter fields,  while in our case  it arose from a standard small-fluctuation analysis around
      a  classical supergravity  solution.  We will comment on  this again  in the next section. 
    \smallskip
      
         One  final rewriting   sheds  further  light on the dual CFT side. 
         We argued earlier    that 
         what  ensures a  weak $3d$  energy-momentum leakage 
          is the scarcity of  the bulk-CFT$_4$  degrees freedom as compared to those on the $3d$  defect. 
    A  measure of  CFT degrees of freedom for   even dimensions  is the familiar Weyl-anomaly coefficient 
    $a$, which for  ${\cal N}=4$
    super Yang Mills with gauge group $U(n)$   is equal to   $n^2$
    [in the normalization
     in which the contribution of a scalar to $a$ is $1/90$]. 
   The analog of $a$ in three dimensions
     is (minus)   the free  energy on   S$^3$, which is   related to 
    the vacuum entanglement entropy across  a spatial circle \cite{Myers:2010tj}\,-\,\cite{Liu:2012eea}. These
    %\cite{Casini:2011kv}\cite{Jafferis:2011zi}
 measures have been    unified by Giombi and Klebanov  in a  generalized free energy \cite{Giombi:2014xxa} 
       defined  (via   dimensional regularization)  for arbitrary  $d$, 
  $\tilde F_d    =  \sin(\pi d/2) \log Z({\rm S}^d)$. For  even dimensions  $ \tilde F_d =  (-)^{d/2} a \pi/2$, while 
   in three dimensions   a holographic calculation  gives 
   $\tilde F_3=  4\pi^2 \langle L_4^2 \rangle  /\kappa_4^2$
     (see for example section 4.2 of \cite{Assel:2012cp}).\,\footnote{The CFT calculation  
    has been performed using  localization in refs.\,\cite{Benvenuti:2011ga}\cite{Nishioka:2011dq}. }
   Combining  everything  we find
    \bea
   \underline{\rm defect} \  \underline{\rm CFT}:\quad \qquad     
    3 (\Delta_{\rm g}-3)   \simeq   m^2_{\rm g}\, L_4^2 
                         \ =  \     {6\pi^3  \tilde F_{4}\over    \widetilde F_{3}}\times     J(\cosh\delta\phi)\,   
   \eea
   where $\Delta_{\rm g}$ is the anomalous dimension of the almost conserved $3d$ energy-momentum tensor.
   This way of expressing the result 
   makes it clear   that the expansion parameter is the  ratio of generalized free energies  
   between the bulk   and the defect CFT.

                                                                                                                                                                                                                                                                                              %%%%%%%%%%%%%%       
      
         \vskip 0.9cm 
           \noindent{\bf\large  6. Bimetric  and double-trace models}                                            
            \vskip 2mm                               
                                                                                                                                                                                                                                                                                              
                                                                                                                                                                                                                                                                                              In ref.\,\cite{Bachas:2017rch}  we analyzed solutions with   a  highly-curved AdS$_5\times$S$^5$ 
                                                                                                                                                                                                                                                                                              throat\,\footnote{The restriction to  AdS$_5\times$S$^5$ 
                                                                                                                                                                                                                                                                                              throats is valid   for  identical bags, or  more generally for  
                                                                                                                                                                                                                                                                                                 bags with  the same  value of the dilaton at the
                                                                                                                                                                                                                                                                                                 two throat entries. This was  implicitly assumed in ref.\,\cite{Bachas:2017rch}.}  
                                                                                                                                                                                                                                                                                                 capped-off at  both of its ends by  bags of  much larger  size. 
  The low-energy theory is in this case a                                                                                                                                                                                                                                                                                               two-graviton  theory. This is a concrete realization   of  the idea of  `Weakly Coupled Worlds'  
  \cite{Damour:2002ws} in which 
  two or more  Universes  endowed with separate  metrics  are coupled  
   through the  mixing of their gravitational fields.    It is well  known
    that massive gravity  can be obtained from bigravity in   a decoupling  limit, and this is also true 
    for our solutions. 
   Before  exhibiting  this decoupling limit, we will 
    first generalize the analysis of \cite{Bachas:2017rch}
                                                                                                                                                                                                                                                                                               from AdS$_5\times$S$^5$  to Janus throats.

                                                                                                                                                                                                                                                                                                   The  manifold M$_6$ now    consists  of a   Janus throat capped-off on  both sides  by 
                                                                                                                                                                                                                                                                                                   two bags,   
                                                                                                                                                                                                                                                                                              $\overline {\rm M}_6$ and $\overline {\rm M}_6^{\ \prime}$.  For economy of notation we introduce  the parameters
                                                                                                                                                                                                                                                                                              $$ v := \int_{\overline {\rm M}_6} \sqrt{g}\, L_4^2  \qquad {\rm and}\qquad 
                                                                                                                                                                                                                                                                                              v^\prime := \int_{\overline {\rm M}_6^\prime}  \sqrt{g}\, L_4^2 \ . 
 $$
                                                                                                                                                                                                                                                                                                     Note that $v$ is just a short-hand for 
                                                                                                                                                                                                                                                                                              the parameter  $V_6 \langle L_4^2\rangle_{\rm bag}  =  \psi_{\rm bag}^{-1/2}$ of
                                                                                                                                                                                                                                                                                               the previous section. 
                                                                                                                                                                                                                                                                                                    Using  the inner product $\langle \psi_1\vert \psi_2\rangle = \int_{{\rm M}_6} \sqrt{g}\, L_4^2\,  \psi_1^* \psi_2$ one finds easily   two orthogonal,  low-lying spin-2 states.   A  massless state 
 with constant wavefunction throughout M$_6$ (which  is normalizable because M$_6$ is now compact), 
                                                                                                                                                                                                                                                                                                  and  a massive state    whose wavefunction is approximately constant in   the bags, 
                                                                                                                                                                                                                                                                                                    \bea\label{xnew}
                      \psi_0(x) \simeq   (v + v^\prime)^{-1/2}\times 
\begin{cases}
\ \sqrt{v^\prime/ v}  \ \ \ {\rm in } \  \  \overline {\rm M}_6  ,\\
- \sqrt{v / v^\prime }  \ \ \ {\rm in } \  \  \overline {\rm M}_6^\prime \ . 
\end{cases}
\eea
                                                                                                                                                                                                                                                                                                   Since  the throat makes a subleading contribution to the inner product,   the above
                                                                                                                                                                                                                                                                                                   wavefunction   is  clearly orthogonal to the constant one, i.e. to the wavefunction 
                                                                                                                                                                                                                                                                                                   of the massless graviton.  This   second  mode  is necessarily   massive  because  $\psi_0$   
                                                                                                                                                                                                                                                                                                   is forced to  vary  inside the Janus throat in order to  extrapolate   between  the above  values
                                                                                                                                                                                                                                                                                                   at the exits.

                                                                                                                                                                                                                                                                                                    One  can now repeat almost verbatim  the calculation of the previous section.  The wavefunction
                                                                                                                                                                                                                                                                                                    in the Janus  throat with the new   boundary conditions reads
                                                                                                                                                                                                                                                                                                    \bea
                                                                                                                                                                                                                                                                                                    \psi_0\  \simeq\  {1\over 2\sqrt{vv^\prime (v -  v^\prime)}}\left[(v^\prime- v)  - (  v^\prime + v )
                                                                                                                                                                                                                                                                                                    {I(x, a)\over I(\infty, a)} 
                                                                                                                                                                                                                                                                                                    \right]\ ,  
                                                                                                                                                                                                                                                                                                    \eea
                                                                                                                                                                                                                                                                                                   where   $I(x,a)$ has been defined in eq.\,\eqref{15}. Inserting the above  wavefunction in  \eqref{17a}, 
                                                                                                                                                                                                                                                                                                   and reexpressing $v$ and $v^\prime$ in terms of  radii and effective couplings   gives
                                                                                                                                                                                                                                                                                                         \bea\label{22a}
      m^2_{\rm g}\, L_4^2 
                         \ =\  {3 \, n^2   \over 16\pi^2  }
                         \left[ {  \kappa_4^2 \over  \langle L_4^2 \rangle_{\rm bag} }  + 
                         {  \kappa_4^{2\, \prime} \over  \langle L_4^2 \rangle_{{\rm bag}^\prime}  }  \right]
                           \times     J(\cosh\delta\phi)\ .  
     \eea \vskip 2mm
                                                                                                                                                                                                                                                                                                   \noindent This agrees with  the result derived  
                                                                                                                                                                                                                                                                                                    in \cite{Bachas:2017rch} for  pure AdS$_5\times$S$^5$ throats for which $J=1$.\,\footnote{ In this reference
                                                                                                                                                                                                                                                                                                   we worked in units of  $L_4=1$ and absorbed  $\langle L_4^2 \rangle$ in the definition of
                                                                                                                                                                                                                                                                                                   $\kappa_4^2$. We thank Thibault Damour for suggesting that we reestablish  explicitly all units.} 
                                                                                                                                                                                                                                                                                                    It also reduces to our formula of   
                                                                                                                                                                                                                                                                                                    the previous section  in  the decoupling limit   $v^\prime \to\infty$, 
                                                                                                                                                                                                                                                                                                    i.e. $\kappa_4^\prime \to 0$   or equivalently $\langle L_4^2 \rangle_{{\rm bag}^\prime} \to \infty$. In this limit 
                                                                                                                                                                                                                                                                                                    the massless graviton has vanishing wavefunction and decouples, whereas  $\psi_0$ is concentrated entirely in the (unprimed)  bag $\overline {\rm M}_6$ and in the throat. 
                                                                                                                                                                                                                                                                                                   
                                                                                                                                                                                                                                                                                                   \smallskip
                                                                                                                                                                                                                                                                                                    
                                                                                                                                                                                                                                                                                                From the perspective of the dual field theory, these bigravity solutions are not 
                                                                                                                                                                                                                                                                                                $4d$ defect CFTs,  but rather $3d$ CFTs of a special kind.  They are   superconformal gauge theories  
                                                                                                                                                                                                                                                                                                based on linear quivers with a 
                                                                                                                                                                                                                                                                                                 low-rank  `weak'    node  \cite{Bachas:2017rch}. Removing this node breaks the quiver into 
                                                                                                                                                                                                                                                                                                 two  disjoint quivers.
                                                                                                                                                                                                                                                                                                 One could in principle integrate out the scarce messenger fields,  thereby generating multitrace couplings between disjoint theories  in the spirit of  \cite{Kiritsis:2006hy}\cite{Aharony:2006hz}.  In contrast with these references, 
                                                                                                                                                                                                                                                                                               the  couplings  are however  non-local  (they are generated by massless messengers)  and exactly scale invariant   (the AdS$_4$  symmetry is manifest).  Conversely, 
                                                                                                                                                                                                                                                                                                integrating  back in the messenger fields  restores the interpretation of  the multitrace couplings 
                                                                                                                                                                                                                                                                                                  in terms of   a classical supergravity background,  and  resolves the conflicts with  string-theory locality
                                                                                                                                                                                                                                                                                                  discussed in refs.\,\cite{Aharony:2001pa}\cite{Aharony:2005sh}. 
                                                                                                                                                                                                                                                                                               
                                                                                                                                                                                                                                                                                                     Similar comments apply to the relation of our models with the transparent boundary conditions of
                                                                                                                                                                                                                                                                                                     \cite{Porrati:2001db}\cite{Duff:2004wh}.  These could conceivably mimic the effects  of 
                                                                                                                                                                                                                                                                                                      the  semi-infinite throats, but they are obscuring the issues of locality and scale invariance. It is nevertheless interesting that they lead to the same parametric dependence of  $m_{\rm g}$ on the effective gravitational coupling $\kappa_4$.

                                                                                                                                                                                                                                                                                %%%%%%%%%%%%%%       
      
         \vskip 0.7cm 
           \noindent{\bf\large  7. Final remarks}                                            
            \vskip 2mm             
                               
             As these top-down embeddings 
           demonstrate, 
             massive AdS$_4$  gravity is   part of the string-theory landscape. 
                 String theory is believed to be a consistent theory,  so we expect the effective $4d$  low-energy
                 gravity  to be free of any  pathologies.  We have seen that   the  
                 effective theory  must break down at the AdS radius $L_4$,  which is comparable to the Kaluza-Klein scale, 
                 a feature that 
                 is related to the scale non-separation problem and could be generic. This still leaves a 
                 range of energies,  $m_{\rm g} < E < L_4^{-1}$,  in which to  try to 
             compare with  effective  actions  such as those  of   refs.\,\cite{deRham:2010kj}\cite{Chamseddine:2018qym}. 
              A technical complication is that string theory is rarely minimal -- the low-energy  theory  would
              have extra  fields in addition to   the massive graviton. 
                 
                        Massive Minkowski  gravity is  harder to embedd and could possibly lie in swampland. 
             One way to see the difficulty is as follows: a key feature of the Karch-Randall model is the existence
             of a local minimum of the AdS scale factor   $L_4$. The existence of a minimum
              seems however  to be  in tension with the holographic $c$-theorem  when the AdS$_4$ fiber is replaced by  Minkowski  
     \cite{Freedman:1999gp}\cite{Girardello:1998pd}\cite{Karch:2000ct}.  It is an interesting question
     whether this obstruction can be somehow relaxed.

                       In a  different direction one can  look for   massive-gravity and bimetric  models in other dimensions
                       and/or with different amounts of supersymmetry.  Many exact  
                       AdS$_D$   solutions with $D>4$ and 
                       half-maximal supersymmetry are  known by now,  for instance
                        \cite{Apruzzi:2015wna} for AdS$_7$,  
                        \cite{DHoker:2016ujz}\cite{DHoker:2017mds}\cite{Apruzzi:2014qva}
                         for AdS$_6$,  and \cite{Gaiotto:2009gz} 
                        for AdS$_5$ (for a review and more references see also \cite{Fazzi:2017trx}). 
                            Some cases can be  a priori excluded.  A prime example is  $6d$ 
                            ${\cal N}=1$, where the  stress tensor belongs to a protected $B$-series multiplet   \cite{Cordova:2016emh}
                            and cannot acquire an anomalous   dimension. Thus  massive AdS$_7$ supergravity is a priori 
                            excluded.\,\footnote{The solutions of \cite{Apruzzi:2015wna} resemble  those
                            studied here --  they are dual to gauge theories based on linear quivers. Anomaly
                            cancellation forces however these latter   to be {\it balanced} quivers. This implies that the 
                              gauge-group rank is a concave function along the quiver,  which 
                              forbids the   low-rank `bridge' nodes of  \cite{Bachas:2017rch}.}
                                The stress tensor multiplet  is  also absolutely protected    for 
                                ${\cal N}=1$   in $5d$, and for more-than-half-maximal supersymmetry
                                (${\cal N}>2$ in $4d$ and  
                                ${\cal N}>4$ in $3d$).  This is consistent with the fact that  there exist   no
                            candidate defect CFTs with so many unbroken  supersymmetries.\,\footnote{The
                               relation of anomalous dimensions with charge leakage  also implies that
                                whenever the stress-tensor multiplet is protected, so are
                                all conserved-current multiplets. Indeed in a  local QFT no charge 
                                can  leak out without carrying also some energy. 
                              Inspection of all superconformal representations in \cite{Cordova:2016emh} 
                              confirms the validity of this assertion. }
                            A situation  with no protection is   ${\cal N} =  2$   AdS$_5$.  It would be interesting to search for
                                 embeddings  of massive AdS supergravity or bigravity in this case. 
                                 It would be even more interesting to search for non-supersymmetric embeddings 
                                 that  allow a Kaluza-Klein cutoff $\gg L_4$ along the lines of \cite{Polchinski:2009ch}.

\vskip 3mm

{\bf Aknowledgement}: We  have benefited from discussions or email exchanges with   Laura Bernard, 
Ali Chamseddine,  Cedric Deffayet,  Eric D'Hoker,  Gregory Gabadadze, 
   Slava Mukhanov, Alessandro Tomassielo and Christoph Uhlemann.  
   C.B. aknowledges the hospitality of  the  
   Mani L. Bhaumik Institute of UCLA  during the last stage of this work.

\vfil\eject

%%%%%%%%%%%%%%%%%%
  %%%%%%%%%%%%%%       
      
         \vskip 0.7cm 
           \noindent{\bf\large   Appendix }                                            
            \vskip 1mm

  Here we collect some formulae on  the exact type-IIB supergravity solutions that we use.
   For more extensive descriptions  of this class of  solutions
  see refs.\,\cite{D'Hoker:2007xy}\cite{D'Hoker:2007xz} and refs.\,\cite{Assel:2011xz}\cite{Assel:2012cj}.

           All   solutions of type-IIB supergravity  preserving  ${\cal N}=4$ AdS$_4$
    supersymmetries have the  form   of a fibration   over a Riemann-surface $\Sigma$, 
     \bea ds^2_{10} \, =\,    L_4^{\,2}   ds^2_{{\rm AdS}_4}  +                                                                                                                                                                                                                                                                                                                                                                                                                                                                                                                                                                                                                                                                                                                                                                                                                                   f^{2} ds^2_{{\rm S}^2 } + \hat f ^{2} ds^2_{\hat {\rm S}^2 } + 4\rho^2 dz d\bar z  
  \eea  
    where the  scale factors of the three (pseudo)spheres 
     depend only on  the coordinate $z = x+i\tau$  of the Riemann surface. For us this latter is
       the  infinite strip, 
    $0\leq \tau \leq \pi/2$.  Solutions based on the annulus do not allow the attachment of semi-infinite Janus throats,  and there 
     are no known  solutions  based on  higher-genus Riemann surfaces. 
     The six-dimensional manifold  S$^2 \times\hat{\rm S}^2 \times\Sigma$ is called M$_6$ in the main text
    (or $\overline {\rm M}_6$ when it is compact). 
     
       The  expressions for the metric factors and  the dilaton, $\phi$,    
    depend on  a pair of  harmonic functions  
     \bea\label{17}
    L_4^{\,8}= 16\, {{ \cal U} \hat{\cal U} \over  W^2} \ ,   \quad 
 f ^{8} = 16 \,h^{8} \,{ \hat {\cal U}  W^2 \over {\cal U}^{\,3}} \ ,
 \quad
\hat f ^{8} = 16 \,\hat h^{8}\, { {\cal U}  W^2 \over \hat {\cal U}^{\,3}} \ , \quad 
  \rho^8 = {{ \cal U} \hat {\cal U}  W^2 \over h^4 \hat h^4} \ ,  
  \quad e^{4\phi} = { \hat {\cal U}    \over {\cal U} }\  ,  
 \eea
where   $h$ and  $\hat h$   are the two harmonic functions,
from which one computes\,\footnote{\,In
  the previous literature  $h, \hat h$ 
are denoted $h_1, h_2$ (and similarly for the associated functions ${\cal U}, \hat {\cal U}$,  the   2-spheres
S$^2$, $\hat{\rm S}^2$, 
and the parameters $\alpha, \hat\alpha$ and $\beta, \hat\beta$).
  We think  that the benefit of making  S-duality  (or  mirror symmetry) manifest outweighs
  the risk of  confusion from  this change of notation . 
} 
 \vskip -5mm
   \bea\label{a1}
\ { \cal U}  = 2 h \hat h |\partial_z h|^2 - h^2 W\  , \quad  \hat { \cal U}  = 2 h \hat h |\partial_z \hat h|^2 - \hat h^2 W\  ,
\quad  {\rm and} \quad 
W = \partial_z \partial_{\bar{z}} (h\hat h) \ .
   \eea  
         The  solutions also have   5-form  and 3-form  fluxes whose explicit form we won't
  need.  
\smallskip                                                                       
                                                                                                                                                                                                               
   After imposing regularity conditions, the  most general solutions with the strip as  base manifold  
  are given by  the 
  following choice for the   harmonic functions:   
                 \bea \label{h1}\nonumber
 h  =   -i\alpha  \sinh (z-\beta )   -   \sum_{a= 1}^N   \gamma_a  \log    \tanh \left(   
 {  {i\pi\over 4} - {z\over 2}  +   { \delta_a \over 2} }   \right) 
   \  + \ c.c.\ , 
\eea            \vskip -6mm
                                             \bea \label{h2} 
\hskip -7mm 
 \hat h  \ =  \   \hat \alpha   \cosh (z- \hat \beta)  -  \sum_{b= 1}^{\hat N} \hat \gamma_b   \log    \tanh \left(   
 { {z\over 2}  -  {\hat \delta_b \over 2} }   \right)     \
  \  + \ c.c. \ . 
\eea  
All  parameters in these expressions are  real. Furthermore the  parameters 
 $\{\alpha , \gamma_a \}$  and $\{\hat\alpha ,  \hat\gamma_b\}$ must all have the same sign 
 which 
can be  chosen positive -- the  harmonic functions are then   positive everywhere inside  the strip.  
The logarithmic singularities on the upper (lower) boundary  of the strip  correspond to    D5-brane (NS5-brane)
sources wrapping 2-sphere cycles. The regions $z\to\pm\infty$, on the other hand, 
correspond to  semi-infinite
D3-brane throats with the  geometry   of the  supersymmetric  Janus solution.

    The pure Janus solution is found by setting  $\gamma_a=\hat\gamma_b =0$. Its  radius is 
    $L_5  =  2  (\alpha \hat \alpha \cosh \delta\phi)^{1/4}$, where 
      $\delta\phi = \beta - \hat \beta$  is the change of the dilaton as $x$ goes from $-\infty$ to $\infty$. 
      By shifting the origin of the $x$ axis one may choose $\beta  = - \hat \beta  = \delta\phi/2$. 
       The combination
      of scale  factors that enters in the  expression \eqref{minimize2} for  the squared mass is  
      \bea  \sqrt{g}\, L_4^4\, \rho^{-2} =
      L_4^4\, f ^2 \hat f ^2 = 16\, h^2 \hat h^2\ . \eea Inserting the Janus  harmonic functions
      gives $16\, h^2 \hat h^2 = (L_5^8/16) \sin^2(2\tau)\, {\cal G}(x)$ where the function   ${\cal G}$ is the one
       defined
      in eq.\,\eqref{G}.  Using finally the $\tau$-independence of $\psi_0$ \cite{Bachas:2011xa}, 
      and integrating over the two 2-spheres and over $\tau$,  leads to the expression \eqref{x}
      for the mass. 
 \smallskip
   
           A typical  `scottish bagpipes' manifolds M$_6$  has $\alpha, \hat \alpha  \ll \gamma_a, \hat \gamma_b$
           and all other parameters $\sim O(1)$.  The   bag $\overline {\rm M}_6$ is 
           obtained in the limit $\alpha , \hat \alpha  = 0$ which truncates away the asymptotic AdS$_5/\mathbb{Z}_2\times$S$^5$ regions of the Janus throats. It can be checked that 
           $x= \pm \infty$ become simple coordinate singularities in this limit, 
           and that $\overline {\rm M}_6$ is  compact 
            \cite{Assel:2011xz}. The product  $\alpha \hat \alpha$ that  controls the  graviton Higgsing 
             is a continuous parameter in supergravity,  but it is  quantized  in string theory where it is related
           to the  D3-brane charges  of the throats (see below).
 
   The detailed shape of the bag depends on the parameters
    $\{ \gamma_a, \hat \gamma_b, \delta_a, \hat \delta_b \} $ which are in one-to-one corerspondence
    with   the NS5-, D5- and D3-brane charges  \cite{Assel:2011xz}
    of the `fat Karch-Randall brane'.   In general the bag has 
   many  different length scales, but a   typical  size   is   $ L_{\rm bag} \sim 
(\sum \gamma_a  \sum \hat\gamma_b)^{1/4}$ where the sum runs over 5-brane singularities
 in some $\delta z \sim O(1)$ region of the strip.  In order to stabilize the bag one 
 needs both NS5-brane and D5-brane
 charges, or else  the dilaton runs away and the solution is singular.

Most  of the details of the bag play no role in the calculation of the graviton mass at  leading order. 
The only relevant parameters are  $\psi_{\rm bag} $  defined in eq.\eqref{C}, and the four combinations
 \bea
\gamma_\pm  := \sum_{a=1}^N  \gamma_a e^{\pm \delta_a}\qquad {\rm and}\qquad 
  \hat \gamma_\pm  := \sum_{b=1}^{\hat N} \hat\gamma_b\, e^{\pm \hat\delta_b}\ . 
 \eea
  To see why let us divide the strip in several regions. For $x \sim O(1)$  we may neglect  
  $\alpha, \hat\alpha$  altogether. This is the bag region  {\footnotesize (I)} of section 5.   For $x$ sufficiently
  large, on the other hand,  we can expand the $\log\tanh$ functions,  and 
  approximate $h$ and $\hat h$  as follows
  \bea\label{expand}
  2h\, \simeq \, -i(\alpha  e^{z- \beta}  - 4\gamma_+  e^{-z})  + c.c.\ , 
  \qquad 2 \hat h \, \simeq \,  (\hat\alpha e^{z-\hat\beta}  +  4 \hat\gamma_+  e^{-z})  + c.c.\ . 
  \eea
   This is the throat region  {\footnotesize (II)} of section 5, where the background is the
   supersymmetric Janus solution. 
   Expressing  \eqref{expand} in terms of   hyperbolic sine and cosine leads to the Janus-parameter   identification
   \bea
   L_5^{(+)} = 2\sqrt{2}  \left[ \alpha\hat\alpha\gamma_+\hat\gamma_+ e^{-(\beta+\hat\beta)} \cosh^2 \delta\phi^{(+)}\right]^{1/8}\quad {\rm and}\qquad    e^{2\delta\phi^{(+)}} =   {\hat\alpha \gamma_+\over \alpha\hat\gamma_+}\,e^{\beta -\hat\beta}\ . 
   \eea
  Similar formulae hold for the parameters $L_5^{(-)},\delta\phi^{(-)}$ of the  
  other  throat, in the region  $x\to -\infty$.  The only  thing to retain from this discussion
  is that for a given  bag (i.e.  given   $\gamma_\pm, \hat\gamma_\pm$) we have enough
   free parameters  to choose  the radii and dilaton jumps of the two asymptotic Janus throats at will.

%%%%%%%%%%%%%%%%%%%%%%%%%%%%
%%%%%%%%%%%%%%%%%%%%%%%%%%%%

\end{document}